\documentclass[12pt]{iopart}

\pdfoutput=1
\usepackage{iopams}
\usepackage{graphicx}
\usepackage{epsfig}
\usepackage[latin1]{inputenc}
\usepackage{amssymb}

\begin{document}

\title{Thermalization through
Hagedorn states -
the importance of multiparticle collisions}
\author{J. Noronha-Hostler$^1$
\footnote{E-mail: hostler@th.physik.uni-frankfurt.de},
Carsten Greiner$^1$
\footnote{E-mail: carsten.greiner@th.physik.uni-frankfurt.de},
Igor Shovkovy$^2$
\footnote{E-mail: Igor.Shovkovy@asu.edu}}

\address{$^1$Institut f\"ur Theoretische Physik, Johann Wolfgang 
Goethe-Universit\"at Frankfurt, Max-von-Laue-Str.1, 
D-60438 Frankfurt am Main, Germany\\
$^2$Department of Applied Sciences and Mathematics, Arizona State University, Mesa, Arizona 85212, USA\\[0.2ex]
}

\begin{abstract}
Quick chemical equilibration times of hadrons within a hadron gas are explained dynamically using Hagedorn states, which drive particles into equilibrium close
to the critical temperature. 
Within this scheme master equations are employed for the chemical equilibration 
of various hadronic particles like (strange) baryon and antibaryons. 
A comparison of the Hagedorn model to
recent lattice results is made and it is found that for both Tc =176 MeV and Tc=196 MeV, the hadrons
can reach chemical equilibrium almost immediately, well before the chemical freeze-out
temperatures found in thermal fits for a hadron gas without Hagedorn states.
\end{abstract}
%

\section{Introduction and Motivation}

As two heavy ions collide color neutral clusters are formed within which the number of particles per cluster increase.  The clusters become so dense and begin to overlap such that it impossible to distinguish quarks from one cluster from that in another i.e. a percolation transition.  The critical density for this is about 
$\epsilon\approx 1\; GeV/fm^3$.
Following the phase transition into Quark Gluon Plasma the interactions are dominated by quarks and gluons.  Through gluon fusions, strange quarks can easily be reproduced. Eventually the QGP cools back into hadrons where the particle yields and ratios are measured.  

If one only considers binary collisions, which react too slowly for strange particles to reach chemical equilibrium within the hadron gas phase, then it is clear that strange particle yields can only be explained through gluon fusion within QGP \cite{Koch:1986ud} and that the hadrons must exist QGP already in full chemical equilibrium \cite{Stock:1999hm}. 
However, multi-mesonic collisions 
$n\pi\leftrightarrow X\bar{X}$
have been demonstrated to reach chemical equilibration
for varoius (strange) antibaryons quickly at SPS  \cite{Greiner}, although they are still not enough to explain the particle yields of exotic antibaryons at the higher energies at RHIC  \cite{Kapusta,Huovinen:2003sa}.  
In order to circumvent such
longer time scales $\sim 10 $ fm/c for a situation of
a nearly baryon-free system with nearly as much antibaryons as baryons, it was then suggested
by Braun-Munzinger, Stachel and Wetterich \cite{BSW} that near Tc there exists
an extra large particle density overpopulated with pions
and kaons, which then drive the baryons/anti-baryons into
equilibrium by exactly such multi-mesonic collisions. But it is not clear how and why this overpopulation of pions and kaons
should appear, and how the subsequent population
of (anti-)baryons would follow in accordance with a standard statistical
hadron model: According to the mass action law the overpopulated
matter of pions will result in an overpopulation of 
(anti-)baryons. For such a large number of
(anti-)baryons it is difficult to get rid of them quickly enough in order to
reach standard hadron equilibrium values before the chemical freeze-out \cite{Greiner:2004vm}.

Rather, understanding the rapid chemical equilibration is possible using Hagedorn states, heavy resonances that drive similar and more multi-hadronic reactions close to $T_c$, as shown in \cite{Greiner:2004vm,Noronha-Hostler:2007fg,NoronhaHostler:2009hp,long}.  
Close to $T_c$ the matter is then a strongly interacting mixture
of standard hadrons and such resonances. Using the Hagedorn states as potential and highly unstable catalysts, the standard hadrons can be populated reactions:
\begin{eqnarray}\label{eqn:decay}
n\pi&\leftrightarrow &HS\leftrightarrow n^{\prime}\pi+X\bar{X}
\end{eqnarray} 
where $X\bar{X}$ can be substituted with  $p\bar{p}$ , $K\bar{K}$ , $\Lambda\bar{\Lambda}$, or $\Omega\bar{\Omega}$.   
The large masses of the decaying Hagedorn states open up the
phase space for multi-particle decays. 

In this note we will compare the particle ratios obtained by using reactions driven by Hagedorn states and those of the experiments at RHIC.  We find that both strange and non-strange particles match the experimental data well within the error bars.  Furthermore, we are able to make estimates for the chemical equilibration time and find that they are very short, which implies that the hadrons can easily reach chemical equilibrium within an expanding, hadronic fireball and that hadrons do not need to be ``born" into chemical equilibrium \cite{Noronha-Hostler:2007fg,NoronhaHostler:2009hp,long}.
Hagedorn states thus provide a microscopic basis for understanding hadronisation
of deconfined matter to all hadronic particles.

Before starting with the details, we emphasize that Hagedorn states have become
quite popular to understand the physics of strongly interacting matter close to the critical temperature:
Hagedorn states have been shown to contribute to the physical description of a hadron gas close to $T_c$.  The inclusion of Hagedorn states leads to a low $\eta/s$ in the hadron gas phase \cite{NoronhaHostler:2008ju}, which nears the string theory bound $\eta/s=1/(4\pi)$ . Calculations of the trace anomaly including Hagedorn states also fits recent lattice results well and correctly describe the minimum of  the speed of sound squared, $c_s^2,$ near the phase transition found on the lattice \cite{NoronhaHostler:2008ju}. Estimates for the bulk viscosity including Hagedorn states in the hadron gas phase indicate that the bulk viscosity, $\zeta/s$, increases near $T_c$ \cite{NoronhaHostler:2008ju}.
We also remark here that Hagedorn states can also explain the phase transition
above the critical temperature and, depending on the
intrinsic parameters, the order of the phase transition
\cite{Zakout:2006zj}.
Finally, it has been shown \cite{NoronhaHostler:2009tz} that Hagedorn states provide a better fit within a thermal model to the hadron yield particle ratios.  Additionally, Hagedorn states provide a mechanism to relate $T_c$ and $T_{chem}$, which then leads to the suggestion that a lower critical temperature could possibly be preferred, according to the thermal fits \cite{NoronhaHostler:2009tz}. 

\section{Setup}

The basis of the Hagedorn spectrum is that there is an exponential mass increase along with a prefactor i.e. the mass spectrum has the form: $f(m)\approx\exp^{m/T_H}$ \cite{Hagedorn:1968jf}. The exponential mass spectrum drives open the phase space, which allows for multi-mesonic decays to dominate close to $T_c$ (we assume $T_H\approx T_c$).    We use the form
\begin{equation}\label{eqn:fitrho}
    \rho=\int_{M_{0}}^{M}\frac{A}{\left[m^2 +m_{r}^2\right]^{\frac{5}{4}}}e^{\frac{m}{T_{H}}}dm.
\end{equation}
where $M_{0}=2$ GeV and $m_{r}^2=0.5$ GeV.  We consider  two different different lattice results for $T_c$: $T_c=196$ MeV \cite{Cheng:2007jq,Bazavov:2009zn} (the corresponding fit to the trace anomaly is then $A=0.5 GeV^{3/2}$, $M=12$ GeV, and $B=\left(340 MeV\right)^4$), which uses an almost physical pion mass, and $T_c=176$ MeV \cite{zodor} (the corresponding fit to the energy density leads to $A=0.1 GeV^{3/2}$, $M=12$ GeV, and $B=\left(300 MeV\right)^4$). Both are shown and discussed \cite{long}. Furthermore, we need to take into account the repulsive interactions and, thus,  use volume corrections  \cite{long,NoronhaHostler:2008ju,Kapusta:1982qd},
which ensure that the our model is thermodynamically consistent. Note that $B$ is a free parameter  based upon the idea of the MIT bag constant.


We need to consider the back reactions of multiple particles combining to form a Hagedorn state in order to preserve detailed balance.  
The rate equations for the Hagedorn resonances $N_{i}$, pions $N_{\pi}$, and the $X\bar{X}$ pair $N_{X\bar{X}}$, respectively, are given by

\begin{eqnarray}\label{eqn:setpiHSBB}
\noindent\dot{N}_{i}&=&\Gamma_{i,\pi}\left[N_{i}^{eq}\sum_{n} B_{i,n}
\left(\frac{N_{\pi}}{N_{\pi}^{eq}}\right)^{n}-N_{i}\right]\nonumber\\
&+&\Gamma_{i,X\bar{X}}\left[ N_{i}^{eq}
\left(\frac{N_{\pi}}{N_{\pi}^{eq}}\right)^{\langle n_{i,x}\rangle} \left(\frac{N_{X\bar{X}}}{N_{X\bar{X}}^{eq}}\right)^2 -N_{i}\right]\nonumber\\
\dot{N}_{\pi }&=&\sum_{i} \Gamma_{i,\pi}  \left[N_{i}\langle n_{i}\rangle-N_{i}^{eq}\sum_{n}
B_{i, n}n\left(\frac{N_{\pi}}{N_{\pi}^{eq}}\right)^{n} \right]\nonumber\\
&+&\sum_{i} \Gamma_{i,X\bar{X}} \langle n_{i,x}\rangle\left[N_{i}-
N_{i}^{eq}
\left(\frac{N_{\pi}}{N_{\pi}^{eq}}\right)^{\langle n_{i,x}\rangle} \left(\frac{N_{X\bar{X}}}{N_{X\bar{X}}^{eq}}\right)^2\right]  \nonumber\\
\dot{N}_{X\bar{X}}&=&\sum_{i}\Gamma_{i,X\bar{X}}\left[ N_{i}- N_{i}^{eq}\left(\frac{N_{\pi}}{N_{\pi}^{eq}}\right)^{\langle n_{i,x}\rangle} \left(\frac{N_{X\bar{X}}}{N_{X\bar{X}}^{eq}}\right)^2\right].
\end{eqnarray}
The decay widths for the $i^{th}$ resonance are $\Gamma_{i,\pi}$ and $\Gamma_{i,X\bar{X}}$, the branching ratio is $B_{i,n}$ (see below), and the average number of pions that each resonance will decay into is $\langle n_{i}\rangle$.  The equilibrium values $N^{eq}$ are both temperature and chemical potential dependent.  However, here we set $\mu_b=0$.
Additionally, a discrete spectrum of Hagedorn states is considered, which is separated into mass bins of 100 MeV. 

The branching ratios, $B_{i,n}$, are the probability that the $i^{th}$ Hagedorn state will decay into $n$ pions where $\sum_{n}B_{i,n}=1$ must always hold. 
We assume the branching ratios follow a Gaussian distribution for the reaction $HS\leftrightarrow n\pi$ 
\begin{equation}
B_{i, n}\approx
\frac{1}{\sigma_{i}\sqrt{2\pi}}e^{-\frac{(n-\langle n_{i}\rangle)^{2}}{2\sigma_{i} ^{2}}},
\end{equation}
which has its peak centered at $\langle n_{i}\rangle\approx 3 - 34$ and the width of the distribution is $\sigma_{i}^2\approx0.8 - 510$ (see \cite{long}). For the average number of pions when a $X\bar{X}$ pair is present, we again refer to the micro-canonical model in \cite{Greiner:2004vm,Liu} and find
\begin{equation}\label{eqn:nfit}
    \langle n_{i,x}\rangle=\left(\frac{2.7}{1.9}\right)\left(0.3+0.4m_i\right)\approx 2-7.
\end{equation}
where $m_i$ is in GeV. In this paper we do not consider a distribution but rather only the average number of pions when a $X\bar{X}$ pair is present.  We  assume that $\langle n_{i,x}\rangle=\langle n_{i,p}\rangle=\langle n_{i,k}\rangle=\langle n_{i,\Lambda}\rangle=\langle n_{i,\Omega}\rangle$ for when a kaon anti-kaon pair, $\Lambda\bar{\Lambda}$, or  $\Omega\bar{\Omega}$ pair is present. 

The decays widths are defined as follows (see \cite{long}):
\begin{eqnarray}\label{decaywidth}
\Gamma_{i}&=&0.15m_{i}-0.0584=250\;\mathrm{MeV\;to}\;1800 \;\mathrm{MeV}\nonumber\\
\Gamma_{i,p\bar{p}}&=&3\;\mathrm{MeV\;to}\;1000 \;\mathrm{MeV}\nonumber\\
\Gamma_{i,K\bar{K}}&=&50\;\mathrm{MeV\;to}\;1700 \;\mathrm{MeV}\nonumber\\
\Gamma_{i,\Lambda\bar{\Lambda}}&=&3\;\mathrm{MeV\;to}\;250 \;\mathrm{MeV}\nonumber\\
\Gamma_{i,\Omega\bar{\Omega}}&=&.01\;\mathrm{MeV\;to}\;4 \;\mathrm{MeV}\nonumber\\
\Gamma_{i,\pi}&=&\Gamma_{i}-\Gamma_{i,X\bar{X}}.
\end{eqnarray}
$\Gamma_{i}$ is a linear fit extrapolated  from the data in \cite{Eidelman:2004wy}. It is then seperated into two parts, one for the reaction $HS\leftrightarrow n\pi$ i.e. $\Gamma_{i,\pi}$ and one for the reaction $HS\leftrightarrow n\pi+X\bar{X}$ i.e.  $\Gamma_{i,X\bar{X}}$.  The decay width $\Gamma_{i,X\bar{X}}$ is found my multiplying $\langle X\rangle$, which is the average X that a Hagedorn state of mass $m$ will decay into, that is found from both microcanonical \cite{Liu,Greiner:2004vm} and canonical models  \cite{long}. 
The large masses open up the phase space for such more special multi-particle decays.
A detailed explanation is found in \cite{long}. 

The equilibrium values are found using a statistical model \cite{StatModel}, which includes 104 light or strange particles from the the PDG \cite{Eidelman:2004wy}. As in \cite{StatModel}, we also consider the effects of feeding for pions, protons, kaons, and lambdas. Additionally, throughout this paper our initial conditions are the various fugacities at $t_0$ (at the point of the phase transition into the hadron gas phase)
$\alpha\equiv\lambda_{\pi}(t_0) \, , \, \beta_{i}\equiv\lambda_{i}(t_0) \, , \mbox{and}\,   \phi\equiv\lambda_{X\bar{X}}(t_0)$
which are chosen by holding the contribution to the total entropy from the Hagedorn states and pions constant i.e. 
$s_{Had}(T_{0},\alpha)V(t_{0})+s_{HS}(T_{0},\beta_{i})V(t_{0})=s_{Had+HS}(T_{0})V(t_{0})=const$
and the corresponding initial condition configurations we choose later can be seen in Tab.\ \ref{tab:IC} (for further discussion see \cite{long}).



\section{Results: Expansion}

In order to include the cooling of the fireball we need to find a relationship between the temperature and the time, i.e., $T(t)$.  To do this we apply a Bjorken expansion for which the total entropy is held constant
\begin{equation}\label{eqn:constrain}
\mathrm{const.}=s(T)V(t)\sim\frac{S_{\pi}}{N_{\pi}}\int \frac{dN_{\pi}}{dy} dy.
\end{equation}
where $s(T)$ is the entropy density of the hadron gas with volume corrections.  The total number of pions in the $5\%$ most central collisions, $\frac{dN_{\pi}}{dy}$, can be found from experimental
results in \cite{Bearden:2004yx}.  Thus, our total pion number is
$\sum_{i}N_{\pi^{i}}=\int_{-0.5}^{0.5} \frac{dN_{\pi}}{dy} dy=874$.
While for a gas of non-interacting Bose gas of massless pions $S_{\pi}/N_{\pi}=3.6$, we do have a mass for a our pions, so we must adjust $S_{\pi}/N_{\pi}$ accordingly.  In \cite{Greiner:1993jn} it was shown that when the pions have a mass the ratio changes and, therefore, the entropy per pion is close to $S_{\pi}/N_{\pi}\approx5.5$, which is what we use here.

The effective volume at mid-rapidity can be parametrized as a function of time.  We do this by using a Bjorken expansion and including accelerating radial flow.  
The volume term is then
\begin{equation}\label{eqn:bjorken}
V(t)=\pi\;ct\left(r_{0}+v_{0}(t-t_{0})+\frac{1}{2}a_{0}(t-t_{0})^2 \right)^2
\end{equation}
where the initial radius is $r_{0}(t_0)=7.1$ fm for $T_H=196$ and the corresponding $t_{0}^{(196)}\approx2 fm/c$. For $T_H=176$ we allow for a longer expansion before the hadron gas phase is reached and, thus, calculate the appropriate $t_0^{(176)}$ from the expansion starting at $T_H=196$, which is $t_0^{(176)}\approx 4 fm/c$. We use $v_0=0.5 $ and $a_0=0.025 $ (see \cite{long}).

Because the volume expansion depends on the entropy and the Hagedorn resonances contribute strongly to the entropy only close to the critical temperature (see \cite{long}),  the effects of the Hagedorn states must be taken into account with calculating the total particle yields otherwise the yields do not increase with the temperature (see \cite{long} for further discussion).  
Therefore, one has to include the potential contribution of the Hagedorn resonances to the pions as in the case of standard hadronic resonances,
e.g. a $ \rho $-meson decays dominantly into
two pions and, thus, accounts for them by a factor two.  
Including Hagedorn states, we arrive at our effective number of pions and $X\bar{X}$ pairs
\begin{eqnarray}\label{eqn:eff}
\tilde{N}_{\pi,X\bar{X}}&=&N_{\pi}+\sum_{i}N_{i}\left[\left(1-\langle X_i\rangle\right)\langle n_{i}\rangle +\langle X_i\rangle\langle n_{i,x}\rangle\right]\nonumber\\
\tilde{N}_{X\bar{X}}&=&N_{X\bar{X}}+\sum_{i}N_{i}\langle X_i\rangle
\end{eqnarray}
because Hagedorn states also contribute strongly to the $X\bar{X}$ pairs close to $T_c$.

\begin{table}
\begin{center}
 \begin{tabular}{|c|c|c|c|}
 \hline
 & & & \\
   & $\alpha=\lambda_{\pi}(t_0)$ & $\beta_{i}=\lambda_i(t_0)$ & $\phi=\lambda_{X\bar{X}}(t_0)$ \\
    & & & \\
 \hline
$IC_1$ & 1 & 1 & 0 \\
$IC_2$ & 1 & 1 & 0.5 \\
$IC_3$ & 1.1 & 0.5 & 0 \\
$IC_4$ & 0.95 & 1.2 & 0 \\
 \hline
 \end{tabular}
 \end{center}
 \caption{Initial condition configurations.}\label{tab:IC}
 \end{table}
Along with the expansion we also must solve these rate equations, Eq. (\ref{eqn:setpiHSBB}), numerically .  We start with various initial conditions as seen in table II and the initial temperature is the respective critical temperature and we stop at $T=110$ MeV. The black solid line in each graph is the chemical equilibrium abundances and the colored lines are the various initial conditions listed in Tab.\ \ref{tab:IC}.  In order to save space, the results are only the expansion  of the $\Lambda$'s and $\Omega$'s results are only shown for $T_H=196$ MeV.  However, $T_H=176$ MeV and further results are shown and discussed in \cite{long}.  However, the end particle ratios are all shown in Fig.\ \ref{fig:summary}. Note that in all the following figures the effective numbers are shown so that the contribution of the Hagedorn states is included.



\begin{figure*}
\begin{minipage}{0.45\linewidth}
\centering
\includegraphics[width=7cm]{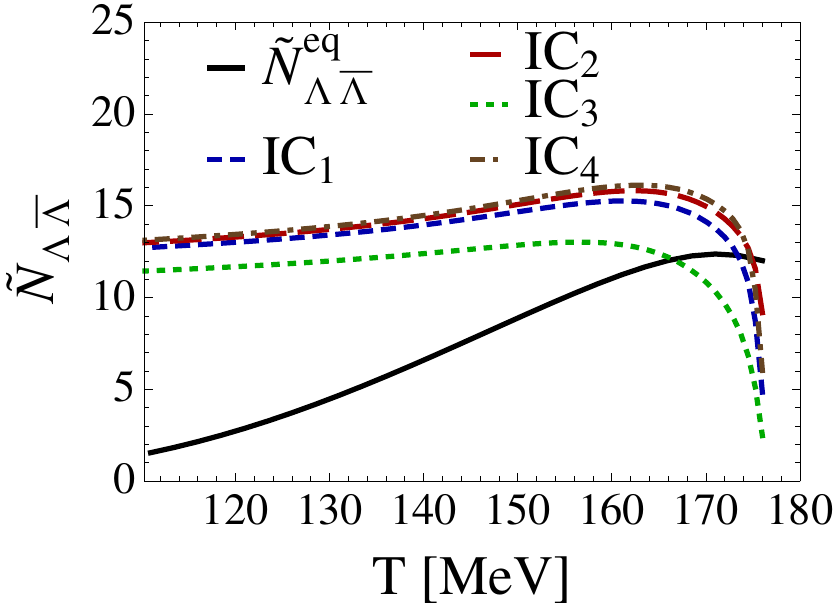} 
\end{minipage}
\hspace{0.5cm}
\begin{minipage}{0.45\linewidth}
\centering
\includegraphics[width=7cm]{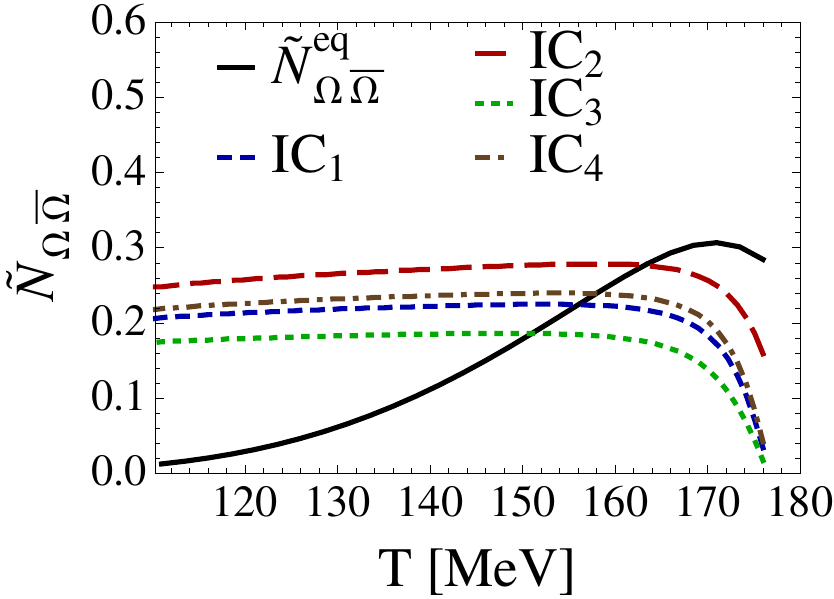} 
\end{minipage}
\caption{Results for $\Lambda$'s and $\Omega$'s with various initial conditions for $T_H=196$ MeV.} \label{fig:LL196}
\end{figure*}

We can also observe the affects of the expansion on the $\Lambda\bar{\Lambda}$ pairs as seen in  Fig.\ \ref{fig:LL196} for the reaction $n\pi\leftrightarrow HS\leftrightarrow n\pi+\Lambda\bar{\Lambda}$. They reach the experimental values almost immediately. One can see that the chemical equilibration time does depend slightly on our choice of $\beta_{i}$, i.e., a larger $\beta_{i}$ means a quicker chemical equilibration time. The one exception is for an underpopulation ($\beta_{i}<1$) of Hagedorn states, which reaches chemical equilibrium by  $T=170$ for $T_H=196$ MeV.  Additionally, when the $\Lambda\bar{\Lambda}$ pairs start at about half their chemical equilibrium values, it only helps the $\Lambda\bar{\Lambda}$ pairs to reach equilibrium at a slightly higher temperature (on the order of a couple of MeV).  

We used our model to investigate the possibility of $\Omega$'s  produced through \cite{Greiner:2004vm}:
\begin{eqnarray}
HS&\leftrightarrow& \Omega\bar{\Omega} +X\nonumber\\
HS\left(sss\bar{q}\bar{q}\bar{q}\right)&\leftrightarrow&\Omega+\bar{B} +X\nonumber\\
HS_B(sss)&\leftrightarrow&\Omega+X.
\label{Omdecay}
\end{eqnarray}
the first we implemented into our model to obtain Fig.\ \ref{fig:LL196} where we are impressively able to  populate the $\Omega\bar{\Omega}$'s.  
However, this scenario is slightly more delicate because $\Gamma_i$ by $50\%$, the total production
of $\Omega $ is not sufficient up to 25 \%,  to meet the 
experimental yield (the other ratios are not significantly affected by the change in $\Gamma_i$).

\begin{figure}
\centering
\includegraphics[width=7cm]{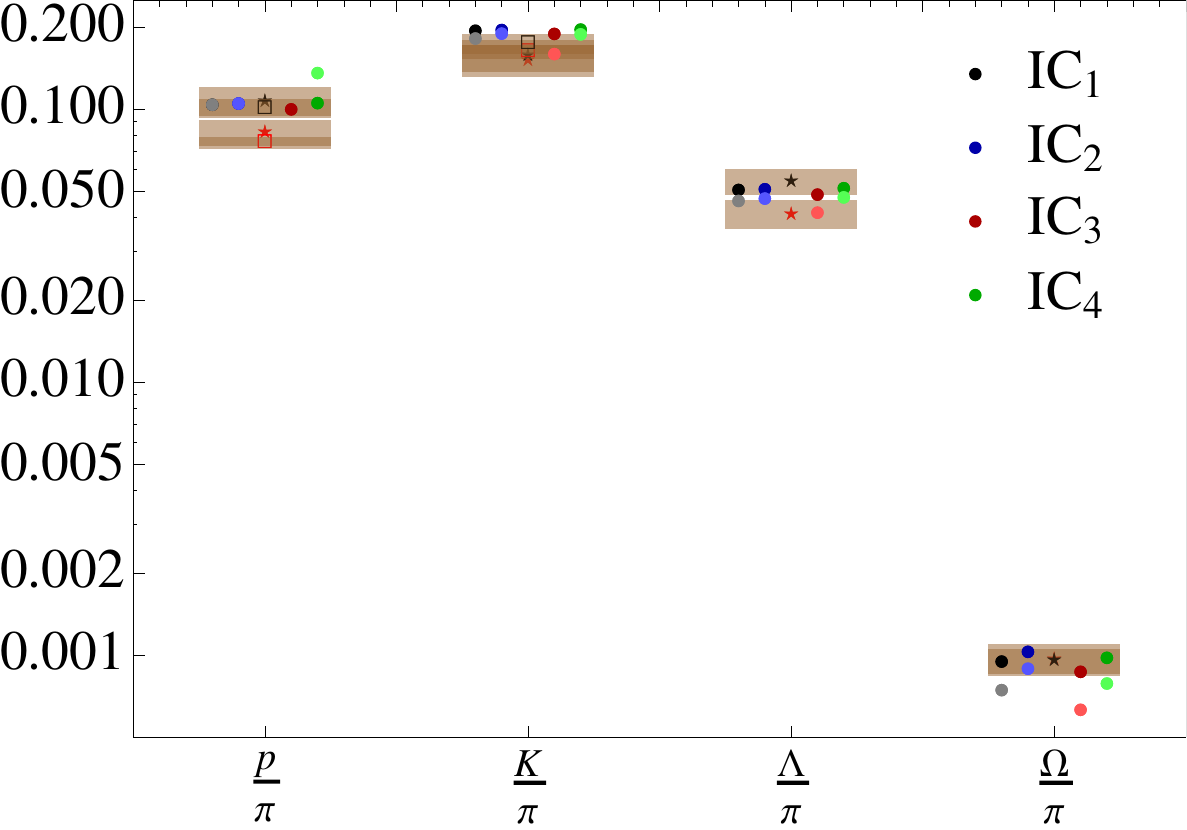} 
\caption{Plot of the various ratios including all initial conditions defined in Tab.\ \ref{tab:IC}.  The points show the ratios at $T=110$ MeV for the various initial conditions (circles are for $T_H=176$ MeV and diamonds are for $T_H=196$ MeV).   The experimental results for STAR and PHENIX are shown by the brown error bars.} \label{fig:summary}
\end{figure}

A summary graph of all our results is shown in Fig.\ \ref{fig:summary}.  The gray error bars cover the range of error for the experimental data points from STAR and PHENIX. 
 The points show the range in values for the initial conditions at a final
expansion point with a temperature $T=110$ MeV. 
We see in our graph that our freezeout results match the experiments well and the initial conditions have little effect on the ratios, which implies that information from the QGP regarding multiplicities is washed out due to the rapid dynamics of Hagedorn states.  
A smaller $\beta_i$ slows the equilibrium time slightly.  However, as seen in Fig.\ \ref{fig:summary} it still fits within the experimental values. Furthermore, in \cite{Noronha-Hostler:2007fg} we showed the the initial condition play almost no roll whatsoever in  $K/\pi^{+}$ and  $(B+\bar{B})/\pi^{+}$.  Thus, strengthening our argument that the dynamics are washed out following the QGP.

\section{Conclusion}
\label{con}

To conclude we have found that the Hagedorn states provide a mechanism for quick chemical equilibration times. Our model gives chemical equilibration times on the order of $\Delta \tau\approx 1-3\frac{fm}{c}$. Furthermore, the particle ratios obtained from decays of Hagedorn states match the experimental values at RHIC very well, which leads to the conclusion that hadrons do not need to be born in chemical equilibrium.    Rather a scenario of hadrons that reach chemical freeze-out shortly after the critical temperature due to multi-mesonic reactions driven by Hagedorn states, is entirely plausible.  We have shown that both strange ($\Lambda$'s and K's) and non-strange ($\pi$'s and $p$'s) hadrons can reach chemical equilibration by $T=160$ MeV. Thus, it would be interesting to implement Hagedorn states into a transport approach such as UrQMD \cite{URQMD}
Moreoever, even multi-strange baryons such as $\Omega$'s can reach chemical equilibrium in such a scenario.  However, unlike the other hadrons, the final abundancy of $\Omega$s is slightly dependent on the decay width of the Hagedorn states (they cannot reach full chemical equilibrium if the decay width is 50\% of its current value) and, thus, should be further investigated using strange and/or baryonic Hagedorn states \cite{Greiner:2004vm}.
Although they are quite exotic, it is still possible that they might occur.  Still, our work indicates that the population and repopulation of potential
Hagedorn states close to phase boundary 
can be the key source for a dynamical understanding of generating and
chemically equilibrating the standard and measured hadrons. 
Hagedorn states thus can provide a microscopic basis for understanding hadronisation
of deconfined matter. 

\section{Acknowledgements}

This work was supported by the Helmholtz International
Center for FAIR within the framework of the
LOEWE program (Landes-Offensive zur Entwicklung
Wissenschaftlich-\"okonomischer Exzellenz) launched by
the State of Hesse.
I.A.S. thanks the members of the Institut f\"ur Theoretische Physik of Johann Wolfgang Goethe--Universit\"at for their hospitality during the final stages of this work. The work of I.A.S. was supported in part by the start-up funds from the Arizona State University.

\end{document}